# Maudlin's Challenge Refuted:

# A Reply to Lewis

R. E. Kastner

University of Maryland, College Park

10 March 2014

ABSTRACT. Lewis has recently argued that Maudlin's contingent absorber experiment remains a significant problem for the Transactional Interpretation (TI). He argues that the only straightforward way to resolve the challenge is by describing the absorbers as offer waves, and asserts that this is a previously unnoticed aspect of the challenge for TI. This argument is refuted in two basic ways: (i) it is noted that the Maudlin experiment cannot be meaningfully recast with absorbers described by quantum states; instead the author replaces it with an ordinary which-way experiment; and (ii) the extant rebuttals to the Maudlin challenge in its original form are not in fact subject to the alleged flaws that Lewis ascribes to them. This paper further seeks to clarify the issues raised in Lewis' presentation concerning the distinction between quantum systems and macroscopic objects in TI. It is noted that the latest, possibilist version of TI (PTI) has no ambiguity concerning macroscopic absorbers. In particular, macroscopic objects are not subject to indeterminate trajectories, since they are continually undergoing collapse. It is concluded that the Maudlin challenge poses no significant problem for the transactional interpretation.

Highlights:

- Lewis' claim that Maudlin experiment remains unresolved is shown to be incorrect.
- Antiparticle offer waves are not absorbers for particle offer waves as claimed.
- Lewis' alternative experiments are ordinary which-way experiments with additional quanta.
- Macroscopic absorbers are well-defined in PTI and do not have indefinite trajectories.
- Probabilities of incipient transactions correspond to the quanta transferred, not to detectors.

1. Introduction and Background

In Lewis (2013), the author argues that the consistency of the Transactional Interpretation of quantum mechanics (TI) continues to face a significant threat from Maudlin's contingent absorber experiment (2002). The author's discussion unfortunately contains quite a few errors and misconceptions. This paper will attempt to clarify some of these issues and to point out that the perception of a problem arises from apparent misunderstandings about TI, especially of its recent development in Kastner (2012a,b) in its 'possibilist' and fully relativistic variant, designated as PTI. In particular, the author's arguments do not take account of the current status of TI as an interpretation that now has clearly defined criteria for what kinds of systems can be considered absorbers and what kinds can be considered offer waves described by quantum states.

It should first be noted that PTI makes no formal revisions to TI, except to extend it into the relativistic domain. Therefore, TI in its original form (Cramer 1986) and PTI both survive the Maudlin challenge. The difference between the original TI and the newer PTI is one of ontology and range: PTI explicitly treats offer waves as pre-spacetime entities, and it includes relativistic interactions as a foundation of the basic transactional process. In particular, the relativistic development in PTI provides for a non-arbitrary definition of absorbers as macroscopic objects that return confirmations with certainty in response to an offer wave. While the possibilist ontology is not crucial for this relativistic extension, it does resolve certain conceptual challenges, such as how to understand the multidimensional nature of multi-particle offer waves. Thus, this paper defends *both* the original TI and PTI, although it is the view of this author that the possibilist ontology is the more conceptually coherent and fruitful approach.

I will now briefly review the Maudlin challenge. Figure 1 illustrates the basic setup of this thought experiment.

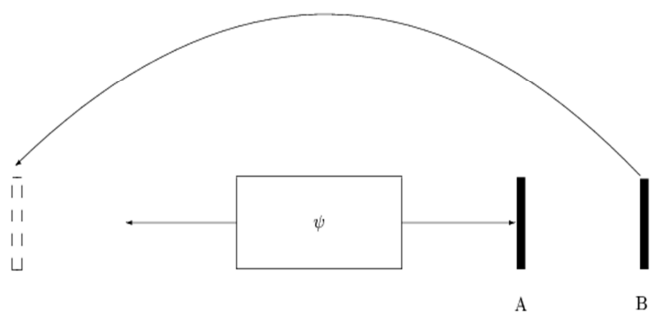

Figure 1: The Maudlin Experiment

As illustrated in Figure 1, a source emits massive (and therefore, Maudlin assumes, slow-moving) particles either to the left or right, in the state $|\Psi\rangle = \frac{1}{\sqrt{2}}[|R\rangle + |L\rangle]$, a superposition of 'rightward' and 'leftward' -propagating states. Offer wave components corresponding to right and left are emitted in both directions, but, in Maudlin's arrangement, only detector A can initially return a confirmation wave (since B is blocked by A). If the particle is not detected at A (meaning that the rightward transaction failed), a light signal is immediately sent to detector B, causing it to swing quickly around to intercept the particle on the left. B then is able to return a confirmation wave of amplitude $\frac{1}{\sqrt{2}}$. At this point, the particle is certain to be detected there, so Maudlin claims that the confirmation wave's amplitude of less than unity is evidence of inconsistency on the part of TI. Another aspect of the alleged inconsistency is the appearance of causal loops: when B returns a confirmation, the emitter receives it at the time of emission, so this seems to imply that B is predetermined to swing to the left. On the other hand, when B does not return a confirmation, the emitter has only a right-hand confirmation at the time of emission, so that seems to imply that B is predetermined *not* to swing to the left.

Maudlin's inconsistency claim has been rebutted by Kastner (2006, 2012b, Chapter 5) and Marchildon (2006) and Boisvert and Marchildon (2013). Lewis takes issue with these rebuttals, and argues that the only 'straightforward' way to resolve the challenge is to treat the absorbers as quantum systems. However, this approach fails to establish the author's claims; specific reasons will be discussed below. In addition, as noted above, TI has recently been developed and elaborated by this author into a fully relativistic interpretation with a possibilist ontology (Kastner (2012a,b, 2014)). This development includes specific physical criteria for macroscopic absorbers, the kind featured in the Maudlin experiment. Much of the author's discussion does not take into account these developments. This leads to a misapplication of the transactional picture, in particular the assignment of offer waves (quantum states) to macroscopic objects that are not offer waves. In addition, the author characterizes the treatment of macroscopic objects in TI as 'murky,' while in fact this issue has been disambiguated in Kastner (2012b, Chapters 6 and 7). In view of apparent ongoing confusion surrounding these issues, I will attempt to clarify that treatment of macroscopic objects here, with further details in Section 3.

In PTI, macroscopic emitters and absorbers are unambiguously defined in terms of their constituent microscopic field currents, which are capable of coupling to other fields. This coupling is characterized by an amplitude, the *coupling amplitude*, which is the amplitude for emission or absorption of another field, as in the emission of a photon by an electron. The value of the coupling amplitude for electrons and photons is the square root of the fine structure constant of quantum electrodynamics (the charge of the electron in natural units). A macroscopic emitter is a system composed of large enough numbers of microscopic emitting currents (e.g. excited electrons) such that the probability of emission of an offer by at least one of the microscopic currents constituting the object is virtually unity.[1] The same definition holds for absorption: a macroscopic absorber, such as a detector in an experiment, is a system composed of large enough numbers of microscopic absorbing currents (e.g. ground state electrons) such that the probability of generation of a confirmation by at least one of the currents is virtually unity. This is discussed in detail in Kastner (2012a, §5), including a representative computation of the probability of generation of a confirmation by at least one of the constituent microscopic currents in a macroscopic sample, showing that it rapidly approaches unity as the number of constituent absorbing currents approaches macroscopic proportions.

The preceding sort of system is the only kind of absorber that instantiates the Maudlin contingent absorber experiment. An absorber cannot be represented by an offer wave undergoing unitary evolution, because the (closer) absorber A in the Maudlin experiment always generates a confirmation and accompanying non-unitary collapse with a specific empirical measurement result, upon which the movement of the more distant absorber (B) is contingent. That is what makes it a contingent absorber experiment.

---

[1] The coupling amplitude leads to a probability when it is taken into account that every amplitude for emission is always accompanied by an amplitude for absorption, and vice versa (no emission or absorption occurs independently in the direct-action picture). This gives us two factors of the coupling amplitude. In fact, in the direct-action picture, this is the natural origin of the QED fine structure constant $\alpha$, which is the square of the coupling amplitude.

2. Lewis alternative experiments are not contingent absorber experiments

As noted above, much of the author's discussion concerns an alleged ambiguity of the applicability of the transactional picture to macroscopic objects. This issue he identifies as a second 'unnoticed' aspect of the Maudlin challenge (the first being the alleged inconsistency of the probability for detection at absorber B). However, PTI is unambiguous about what counts as a macroscopic absorber (and is decidedly not an offer wave) and what counts as a microscopic object (which could be described by an offer wave or at least as a bound state describable to a good approximation by a quantum state). Perhaps due to a misunderstanding of this work, the author presents two 'alternative' versions of the Maudlin experiment that misapply the transactional picture. In one of these, a quantum system, an antiparticle, is mis-identified as an absorber; while in the other, allegedly macroscopic absorbers are assumed to be describable as offer waves. Rather than elucidate an 'overlooked' aspect of the Maudlin challenge as the author claims, these scenarios unfortunately create confusion both about the Maudlin experiment and about TI.

Specifically, in eqs . (1) and (2) the author proposes an antiparticle offer wave as an absorber for a particle offer wave. (I assume that examples of the proposed particles would be an electron and a positron.) However, an antiparticle offer wave cannot be an absorber for a particle offer wave. An offer wave of a given field, such as the Fermi-Dirac field for fermions, does not couple to other offer waves of the same field, and therefore does not constitute an absorber for offer waves of the same field. Of course, a fermion offer wave may be *indirectly* detected by another fermion, for example in the case of a bound electron being excited by another approaching electron due to the transactional exchange of a photon. But this possibility at the micro-level does not allow the antiparticle in this experiment to be viewed even as an effective absorber, since it is participating in unitary evolution (this is discussed further below).

In addition, it is not the case that a confirmation wave is sent "from the point at which the particle and antiparticle annihilate" (Lewis 2013, discussion under eqn. (2)). A confirmation wave is always generated by a specific entity, and again the antiparticle is not an absorber for the particle. Ordinary relativistic quantum theory treats the interaction between the particle and antiparticle offer waves as a scattering process, as does TI, especially in this author's updated version (Kastner 2012b, §6.3). A process in which an incoming particle and antiparticle offer undergo scattering has a well-defined amplitude (less than unity) for annihilation, in which the outgoing states are two photons. Even if we ignore that and pretend that the amplitude for annihilation is unity for each path, there is certainly no confirmation sent from that interaction stage. Those outgoing photons are just offer waves. The only information one has concerning 'which way' both particles went is by way of photon confirmations sent from fixed, macroscopic photon detectors that the author doesn't include in the experiment.

But, as alluded to above, the more important point is that the antiparticle offer wave is part of a composite particle/antiparticle offer wave with an overall amplitude of $1/\sqrt{2}$ for both paths. This dictates that the probability for confirmation to be sent from a putative absorbing antiparticle that is *itself* described by either term (even if it could) would be (at most) only ½. In contrast, in a contingent absorber experiment, a confirmation is *certain* to be sent from the relevant detector, which generates confirmations with a probability of unity. The status of the macroscopic detector as an object that exists in spacetime with a probability of unity and generates confirmations with a probability of unity (in contrast to the antiparticle described by an amplitude of $1/\sqrt{2}$) is further clarified in the next section.

Thus, the author's proposed experiment fails to capture the crucial contingent absorber features of the Maudlin challenge. The antiparticle is not really an absorber, and the absorbers actually generating the confirmations with certainty for each term of the superposition are fixed. The antiparticle, as a unitarily evolving quantum system, becomes correlated with an offer wave in a superposition of two ways and thereby acts as a microscopic pointer. But that pointer does not yield collapse, which is the hallmark of absorption; instead it must be collapsed by another absorber that will generate confirmation with certainty (i.e., a macroscopic absorber). So we end up with an ordinary 'which way' experiment, in which we are just measuring which way both particles went. However, the author's argument does show that if the Maudlin experiment cannot actually be carried out with a truly macroscopic absorber, i.e., one that always generates confirmations with certainty, then it poses no challenge whatsoever for TI.

We see therefore that the translation of the Maudlin challenge into a 'quantum mechanical' form makes it vacuous, rather than offering a 'straightforward solution' to the problem as the author contends. Of course, there is no challenge, and the situation is perfectly describable in TI, but this is not a contingent absorber experiment; it is just a which-way experiment with an extra particle along 'for the ride'. Nor does this argument demonstrate that nullifying the experiment is the only way to resolve the puzzle presented by the experiment in its non-nullified, original form. That is, Lewis' claim that the only way to present an 'adequate realist account' of the Maudlin experiment is to describe absorbers by quantum states is false, as will be shown in §4 and §5. The same basic point applies to the second alternative version in which quantum states representing offer waves are erroneously assigned to macroscopic absorbers. This second version will be discussed further below.

3. Macroscopic objects well-defined in PTI

It has been objected in the past that a 'macroscopic object' is an allegedly ill-defined or problematic concept in TI. While this may once have been the case, it is long past the time that this criticism was a fair one. The concept of an absorber as a macroscopic object is discussed in detail in Kastner (2012a, section 5). As noted above, absorbers are defined in terms of coupling amplitudes, which are naturally interpreted in a direct-action theory (such as the Wheeler-Feynman (1945, 1949) and Davies (1971, 1972) theories upon which TI is based) as the amplitudes for both the emission of offer waves and the generation of confirmations.[2] A macroscopic absorber, such as Maudlin's detectors A and B, is a system in which the probability for the generation of a confirmation by at least any one of its constituent coupling microscopic currents approaches unity. A macroscopic absorber is virtually guaranteed to generate a confirmation due to its composition by large numbers of microscopic currents capable of coupling to a given offer wave. Similarly, a macroscopic emitter (such as a laser) is virtually guaranteed to generate offer waves.

Upon the actualization of a transaction, both the emitter and the absorber are localized and are therefore macroscopic, spacetime events. Therefore, such an object is never accurately described by a quantum state (offer wave), since it is continually being localized due to ongoing emission or absorption, and collapse. In contrast, offer waves are nonlocal objects, since they are field excitations.[3] This is how PTI naturally demonstrates the emergence of a classical, macroscopic realm from a domain of quantum possibilities. While the account of the emergence of a macroscopic absorber is not deterministic, it is quantitatively defined, unambiguous, and wholly realist. Since it is the macroscopic, collapsing kind of absorber that creates the Maudlin challenge in the first place, the situation discussed in equations (4) and (5) in which macroscopic absorbers are erroneously described by offer waves, fails both as a version of Maudlin's challenge and as an application of TI.

---

[2] Feynman himself noted that the coupling amplitude for QED is the amplitude for a real photon to be emitted or absorbed (Feynman 1985). In a direct-action picture such as that of Wheeler-Feynman (1945, 1949) or Davies (1971, 1972), upon which TI is based, this is the amplitude for emission of an offer wave *or* for generation of a confirmation wave.

The author unfortunately creates further confusion about TI's treatment of macroscopic objects in the following statement: "The worry about incorporating macroscopic objects into a TI-style analysis is that it opens the door for macroscopic objects to have indeterminate trajectories." For one thing, it is not clear what the author means by "TI-style." If by this he means the transactional interpretation in its up-to-date form (Kastner 2012a,b), there is in fact no such worry, and to suggest that there is misrepresents the interpretation. If, on the other hand, the author has in mind another version of TI with worries not applicable to the version in Kastner (2012a,b), he needs to make that clear in order to avoid unnecessary confusion.

In any case, the alleged worry about macroscopic objects does not apply to TI in its latest form. As discussed in Kastner (2013a,b), it is never appropriate to describe macroscopic systems – by which we mean systems composed of enough microscopic currents to generate either offers or confirmations with certainty -- by quantum states. This is because quantum states only apply to systems that are either excited states of fields (offer waves) or quantum bound states with very small amplitudes (i.e. of the order of the fine structure constant) of emitting offers or generating confirmations. In contrast, as noted above, macroscopic objects are always localized by ongoing actualized transactions. That is how the existence of macroscopic objects is explained in PTI – they are continually undergoing collapse that localizes them in spacetime, so they can't have indeterminate spacetime trajectories.[4]

---

[3] This may seem to indicate a 'preferred observable' in that conserved quantities such as energy and momentum are actualized and transferred in transactions. This is a natural reflection of the fact that position is not an observable in relativistic quantum theory, which is the more accurate theory. Nonrelativistic quantum theory, which treats position as an observable, is only an approximation, and should never be viewed as a free-standing theory. It is through actualized transactions that objects become localized – spacetime is just the arena of actualized transactions, and spacetime events are actualized emission and absorption events. We never really 'observe' position. We simply participate in transactions that transfer conserved physical quantities between our sense organs and other objects.

[4] One can assign an 'effective quantum state' to a macroscopic object, but only as a kind of correspondence principle, e.g., by ascribing to it a center-of-mass position state with so little uncertainty that it can be regarded as having a determinate position. However, in general, macroscopic objects are conglomerates of actualized transactions, and therefore are not quantum objects. In any case, such a system is certainly not an offer wave.

Thus, neither the antiparticle nor the putative 'quantum absorbers' function as real absorbers in the author's proposed alternative version of the Maudlin experiment. An 'absorber' in TI is an object that generates confirmations with certainty, and that always results in collapse, a nonunitary process. However, the author's revision of the experiment with quantum states for the detectors A and B describes the entire process as unitary; thus the 'absorbers' are not really absorbers. The whole point of the Maudlin challenge is as a contingent absorber experiment; absorber *B* is made accessible to the left-hand offer wave contingent on the failure of the transaction with *A*. But in the author's versions, the generation of a confirmation wave is not contingent on anything, since all the relevant detectors are available to the offer waves at all times. These are the photon detectors overlooked in the author's first version, and unspecified other fixed detectors bringing about the 'destruction' of the particles putatively comprising the detectors A and B in his second version. The alternative experiments are just ordinary which-way experiments involving additional quantum systems that could go either way along with original particle. So the author fails to construct a 'quantum' version of the Maudlin experiment, and his arguments based on that premise fail as well.

4. Criticism of Marchildon rebuttal fails

The author claims that the only 'straightforward' solution to the Maudlin experiment is by essentially treating all the objects as offer waves – which means, in view of our arguments above, to nullify the Maudlin experiment by changing it into an ordinary which-way experiment. Does this mean that the rebuttals of Kastner and Marchildon fail to resolve the challenge? No, because those rebuttals have apparently been misunderstood.

Marchildon's solution (2006) is elaborated in Boisvert and Marchildon (2013). To review, Marchildon notes that in the original conception of the Wheeler-Feynman direct-action theory, there is always absorption somewhere for any emitted offer wave; the universe is a 'light-tight box'.[5] Thus Marchildon includes a remote, background absorber C, which always returns confirmations from the left side. (The left side is where the closer, moveable absorber B moves if the right-hand detector A does not fire.) This means that there are always confirmations from both the left and the right, and there are always two incipient transactions with probabilistic

---

[5] Recall that the Wheeler-Feynman (1945, 1949) and Davies (1971, 1972) theories of the direct-action picture, upon which TI is based, are empirically equivalent to standard theories of fields if the universe is a 'light-tight box' or if other equivalent boundary conditions obtain -- an alternative BC is explored in Cramer (1983).

weights of ½, correctly reflecting the observed frequencies of detection on the right and the left. When detector A does not fire, it means that the particle is emitted to the left and is intercepted by B. The author presents Marchildon's solution and comments:

"Marchildon's account retains the inconsistency …. regarding absorber B; absorber B definitely receives the particle if it has a probability of 1/2 of doing so. Furthermore, his account adds a new inconsistency regarding absorber C; absorber C definitely does not receive the particle when it has a probability of 1/2 of doing so. The contradiction at the heart of Maudlin's challenge remains." (Lewis 2013)

No, the alleged contradiction does not remain. The mistake in the above argument is in ascribing the probabilities to detectors. The probabilities apply to the realization of the properties of the quantum being transferred, not to a particular detector. This is so because *it is the confirmed field itself*, not emitters or absorbers, which instantiates the physical probabilities. So for example, the particle offer wave component that reaches C is characterized by a particular energy and momentum. It does not matter whether that actualized particle is ultimately absorbed by the movable detector B or at C; it is the energy and momentum of the particle that is actualized, not 'detection at B' or 'detection at C.' Regardless of which detector ends up receiving the energy, *the same* particle*, with the same energy and momentum, is detected*. And it is detected (actualized) on the left with a probability of ½, meaning that half the time a particle with leftward momentum is actualized and half the time a particle with rightward momentum is actualized. Thus, there is no inconsistency.

The mistake in the author's argument is in attributing probabilities to specific detectors, rather than to the entities actually described by the probabilities – the quanta being emitted and absorbed. This misconception is understandable, since standard applications of quantum theory routinely attribute the probabilities to detection sites for pragmatic purposes. But the probabilities themselves actually apply to *the transfer of the physically conserved quantities described by the quantum states* (energy, momentum, angular momentum, etc.) So clearly, including a distant absorber resolves any apparent inconsistency concerning the physical meaning of the weights of the incipient transactions, which function as probabilities of actualization of those transactions. Whenever the rightward transaction is not actualized, the leftward one is, and they both have a probability of ½, and those are expressed in the observed frequencies of the respective outcomes.

## 5. Criticism of Kastner rebuttal fails

It is clear from the above that Marchildon's solution is a very straightforward one. However, it does depend explicitly on the 'light tight box' condition. In an effort to address the Maudlin challenge on its own terms (i.e., incomplete absorption)[6], I have pointed out (Kastner 2012b, Chapter 5) that even in this case, Maudlin's experiment presents no more inconsistency problem for TI than Wheeler's 'delayed choice experiment'[7] does for standard quantum theory, provided one ascribes a particular type of behavior to the photon based on the experimental configuration. In fact this is commonly done in the literature. For example, Peruzzo et al (2012), present their interesting variation on the delayed choice experiment and note that the delayed choice determines whether the photon will previously have demonstrated wavelike or particle-like behavior. In this context, 'wavelike' means both-routes travel, while 'particle-like' means which-route travel. Indeed, Peruzzo et al take their experiment to be measuring this behavior on the part of the photon (cf. their discussion under eqn. (1)).

Regarding Wheeler's own ontological interpretation of the delayed choice experiment, he first comments that : "…one decides whether the photon 'shall have come by one route or by both routes' after it has already done its travel." (Wheeler 1981, p. 183). But then he qualifies that formulation by saying that "In actuality it is wrong to talk of the route of the photon," because (in Copenhagen fashion) he believes that one cannot speak about what happens apart from 'registered phenomena.' Nevertheless, Wheeler asserts that the delayed choice 'reaches into the past' to dictate what the overall phenomenon will be.

Now, if one *is* allowed to consider the photon as engaging in some definite kind of behavior (either wave or particle), then a Maudlin-type inconsistency arises for the delayed choice experiment, including apparent causal loops (this point is reviewed below). On the other hand, if Wheeler's comments about 'reaching into the past' are intended to assert that the past is indeterminate prior to the delayed choice, in a kind of growing universe picture, then there is no issue with causal loops nor any inconsistency with the probabilities. Indeed, the indeterminate past is part of the PTI 'growing universe' picture (Kastner 2012, Chapter 8) , so that also removes the alleged inconsistency facing the transactional picture in connection with the Maudlin experiment.

---

[6] Since direct-action theories require full absorption, or an equivalent boundary condition (such as a perfectly reflecting t=0 condition) for empirical equivalence with standard theories of radiation, any scenario lacking full absorption is speculative, and responses to the Maudlin challenge on those terms necessarily inherit that speculative quality.

[7] Wheeler reviews this proposal in Wheeler (1981).

To review these points, let us first consider the case in which we can consider the photon to be following either a 'which way' (particlelike) route or a 'both ways'(wavelike) route. The causal loop consists in the fact that the choice of an experimenter, at a time $t_2$ whether to do a 'which way' or 'both ways' measurement exerts some causal influence on the past of a photon at a time $t_1<t_2$; that is, it affects whether the photon *previously* 'went through both slits' or 'went through only one slit'. In the usual block world formulation, the future choice at $t_2$ must therefore *already be determined* according to what the photon does at the slits at $t_1$. This gives rise to an apparent causal loop of the same kind as the Maudlin challenge: given whatever the photon does at $t_1$, the experimenter *must* choose the corresponding measurement at $t_2$. If one replaces the experimenter's choice by a quantum coin flip with probabilities of ½ for each option, then the probabilities applying to the quantum coin flip are inconsistent.

To see this, suppose 'heads' instructs us to 'measure which slit' and 'tails' instructs us to 'measure both slits and get interference.' Each time the coin flips (according to the block world picture), its outcome *must already be decided* by whatever the photon did before the coin flip. If the photon went through both slits (i.e., exhibited wavelike behavior), then it is certain that the coin will come up 'tails,' yet the quantum probability for that outcome is only ½. This is exactly the same alleged inconsistency that the Maudlin experiment presents for TI. To the extent that the consistency problem can be deflected by Wheeler's argument that "it is wrong to think of the past as 'already existing' in all detail," (Wheeler 1981, p. 194), so can the consistency challenge of the Maudlin experiment be deflected in PTI, since according to PTI past events also do not exist determinately in spacetime unless they are brought about in actualized transactions.

Thus the argument in Kastner (2012b, Chapter 5) points out that most formulations of the delayed choice experiment involve at least a tacit assumption that the photon either 'went by both routes' or 'went by one or the other route', and that these are therefore presented with the same alleged consistency problems as the Maudlin experiment seems to present for TI. However, this is not viewed as a severe problem in the literature (indeed, it does not even seem to have been noticed); and therefore the elimination of an untenable double standard dictates that it cannot be a severe problem for TI either. Lewis' response is to note that standard quantum theory has unresolved conceptual problems, such as the measurement problem, and argues that it should not be considered reassuring to be told that TI is no more threatened by the Maudlin challenge than standard quantum theory is by the delayed choice experiment (where standard quantum theory is interpreted as allowing 'which path' or 'both paths' talk). In this regard, the author cites other interpretations not subject to the Maudlin challenge that he claims solve the measurement problem.[8] But in fact it is far from settled that the other approaches cited by the author really do solve the measurement problem or satisfactorily explain the Born Rule. Meanwhile, I have argued (Kastner 2012, Chapter 3) that the transactional picture readily accomplishes both.

In any case, the point of the rebuttal reviewed above was to illuminate an apparent double standard lurking in judgments that the Maudlin challenge is fatal for TI. Rather than eliminate the double standard by acknowledging that in fact neither apparent problem is a severe one for standard quantum theory or for TI, the author attempts to minimize the double standard by arguing that *both* apparent problems are severe. Specifically, in arguing that the delayed choice experiment presents a severe problem for standard quantum theory, the author claims that this was a motivation for John Cramer in developing TI, and invokes Cramer's discussion (1986, p. 671) of that experiment. But this is misleading. In the cited section, Cramer was discussing how TI can provide an observer-free account of measurement, and pointing out that including the response of absorbers resolves the alleged involvement of a conscious observer in determining whether a photon retroactively went through one or both slits in the delayed choice experiment. Thus, clearly Cramer's intent was to eliminate the conscious observer and to solve the measurement problem in physical terms, not to claim that the delayed choice experiment specifically posed a severe consistency problem for quantum theory.

---

[8] The other interpretations claimed by the author to solve the measurement problem are the Bohmian theory, Everettian (MWI) approaches, and the GRW theory. However, all these interpretations suffer from serious deficiencies not faced by TI, such as difficulty explaining the Born Rule, or what causes the frankly *ad hoc* 'collapses' of the GRW approach. (Also, the status of GRW approaches as comprehensive interpretations capable of dealing with both nonrelativistic and relativistic realms is far from clear, and their alleged 'resolution' of the measurement problem is bought at the high price of changing quantum theory.) It is not even clear that the Bohmian theory does solve the measurement problem (see, e.g., Brown and Wallace 2005), so the author cannot take this for granted. In addition there is great difficulty in extending the Bohmian theory to the relativistic domain, while TI applies seamlessly to both the nonrelativistic and relativistic domains as shown in Kastner (2012a,b, 2014). Regarding MWI, Kent (1990) argues that "no plausible set of axioms exists for an MWI that describes known physics." He extends his critique in a more recent publication (2010), concluding that "...no known version of the theory (unadorned by extra ad hoc postulates) can account for the appearance of probabilities and explain why the theory it was meant to replace, Copenhagen quantum theory, appears to be confirmed, or more generally why our evolutionary history appears to be Born-rule typical" (Kent 2010) So it is far from established that the alternative interpretations cited by the author succeed in resolving the measurement problem or other challenges in understanding quantum theory.

Apart from this apparent misstatement by the author of Cramer's intent regarding the delayed choice experiment, the basic point is that if one views aspects of the past as indeterminate prior to the delayed choice, which then 'reaches back into the past' to bring about a determinate situation, the apparent inconsistency is removed. This remedy applies just as well to the possible confirmation wave from B in the context of the Maudlin experiment. So there is no need to attempt to remedy the double standard in the negative way chosen by the author; i.e., by arguing both (1) that the delayed choice experiment poses a serious consistency threat to quantum theory *and* (2) that the Maudlin challenge poses a serious consistency threat to TI. In fact, neither is the case.

A further significant inaccuracy in the author's discussion of Cramer (1986) needs to be noted here. Specifically, he claims that TI can be viewed as similar to the Copenhagen Interpretation (CI). However, these two interpretations – TI and CI -- could scarcely be more different. Cramer (1986) takes great pains to contrast TI with CI, and presents TI as a more effective, realist approach which clearly defines the measurement process rather than taking macroscopic objects as primitive and relying on the intentions and behaviors of conscious observers to define measurement, as in CI. Even if the original TI appears to have taken absorbers as primitive (which is debatable) and thereby to invite comparison with CI in terms of the macroscopic/microscopic divide, that is certainly no longer the case with the updated versions of TI in the literature (Kastner 2012a,b) . That work presents clear physical, observer-free criteria for the macroscopic/microscopic boundary in the transactional picture. In view of the explicit opposition to CI in Cramer's original 1986 proposal, and recent developments of TI, all of which are clearly aimed at providing a specific physical account of collapse as opposed to CI's appeal to observers in a primitively classical world, this author finds it hard to understand how TI could be reasonably considered to have anything of significance in common with the Copenhagen interpretation.

6. Conclusion: Maudlin challenge has been decisively refuted

It has been shown herein that Lewis' arguments fail to show that the Maudlin experiment presents an ongoing significant problem for the transactional picture, either in its original from (TI) or its recently developed form (PTI). In several instances the author's arguments mischaracterize and/or misapply the transactional picture, and his alternative versions of the Maudlin experiment are not contingent absorber experiments. All they do is to nullify the experiment rather than render it resolvable through a 'straightforward solution' as claimed. In fact, the simplest and most straightforward solution is that of Marchildon (2006): in the direct-action theory upon which TI is based, there is a fully absorbing background universe or alternative boundary condition (cf. Cramer 1983) that results in a full set of confirmation waves at the emitter. In this case, there are two well-defined incipient transactions with equal weights, and those correspond to probabilities expressed in the frequencies of the actualizations of leftward- and rightward-directed particles. The probabilities do not apply to specific detectors, but rather the particles themselves, which are the entities described by the weights of the incipient transactions (as opposed to the detectors B and C).

Alternatively, if one wants a more general (but necessarily speculative) approach that does not assume complete absorption, one can opt for the observation in Kastner (2012b, Chapter 5) that in a growing universe the past may be indeterminate without any inconsistency. Moreover, Wheeler's delayed choice experiment gives rise to a similar apparent inconsistency of probabilities in a 'block world' picture in which one assumes either wavelike or particle-like behavior for the unobserved photon while it is in the apparatus. These are no more than interesting puzzles for standard quantum mechanics, not fatal problems (to the extent that they have even been noticed). Thus there is a double standard involved in treating the Maudlin challenge (in its original but speculative incomplete absorber context) as a serious problem for TI.

In addition, an attempt has been made herein to clarify issues that have been subject to apparent confusion, especially the treatment of macroscopic objects in the latest version of the transactional picture (PTI). Macroscopic emitters and absorbers are unambiguously defined in PTI. These objects cannot accurately be described by quantum states, since they are neither excited field states nor quantum bound states. Rather, they are conglomerates of ongoing actualized transactions, which *physically* (not epistemically) localize them as classical, persistent spacetime events. Thus such objects cannot have indeterminate trajectories as suggested by the author, nor can they be accurately described as offer waves. In addition, extant rebuttals of the Maudlin challenge by Marchildon and Kastner have been shown to be effective when correctly understood. Thus Maudlin's challenge has been effectively disarmed, and no longer presents a significant problem for the transactional interpretation.

Finally, the present author recognizes that some researchers view the transactional picture as speculative or controversial. However, it should be kept in mind that even currently 'mainstream' interpretations, such as the Bohmian theory or Everettian (Many-Worlds) approaches, are not without significant controversy, and contain speculative elements (as briefly reviewed in footnote #7). Moreover, the Bohmian approach was considered marginal for decades before it became viewed as mainstream. It is probably safe to say that the task of interpreting quantum theory has never been without controversy. Newer approaches are often initially viewed as 'speculative,' but that does not mean that they are undeserving of serious consideration. Science advances by careful consideration and development of initially speculative ideas; some of these may ultimately fail, but many succeed. Boltzmann's atomic hypothesis, considered at the time by Mach to be speculative in the extreme, is certainly an example of the latter.
Acknowledgments.
I would like to thank two anonymous referees for their valuable suggestions for the improvement of the manuscript.